# Intrinsic Instability of Aberration-Corrected Electron Microscopes

S.M. Schramm[1], S.J. van der Molen[1], and R.M. Tromp[1,2]

[1] Leiden University, Kamerlingh Onnes Laboratorium,
P.O. Box 9504, NL-2300 RA Leiden, The Netherlands
[2] IBM T.J. Watson Research Center, 1101 Kitchawan Road, P.O. Box 218,
Yorktown Heights, NY 10598, USA

*Aberration-corrected microscopes with sub-atomic resolution will impact broad areas of science and technology. However, the experimentally observed lifetime of the corrected state is just a few minutes. Here we show that the corrected state is intrinsically unstable; the higher its quality, the more unstable it is. Analyzing the Contrast Transfer Function near optimum correction, we define an 'instability budget' which allows a rational trade-off between resolution and stability. Unless control systems are developed to overcome these challenges, intrinsic instability poses a fundamental limit to the resolution practically achievable in the electron microscope.*

Correction of spherical and chromatic aberrations of the electron microscope constitutes one of the most far-reaching breakthroughs in electron optics in the last 20 years[1]. Now, Transmission Electron Microscopy (TEM) with 50 pm resolution provides a detailed view of carbon atoms in a single sheet of graphene[2]. The resolution of Low Energy Electron Microscopy (LEEM) has improved[3] from 5-10 nm to below 2 nm, with a theoretical limit of ~ 1 nm, opening up new possibilities for the dynamic imaging of surfaces, interfaces and thin films, including domain boundaries and domain walls, as well as nanometer-scale organic and biological materials. Photo Electron Emission Microscopy (PEEM), uniquely suited for elemental, chemical, electronic, and magnetic imaging, has achieved[4] ~5 nm resolution, with a further factor 2 improvement still possible. Aberration correction has also been applied to Scanning Electron Microscopy (SEM)[5] and Focused Ion Beam (FIB)[6] systems, and is being considered for applications in semiconductor electron beam lithography and inspection tools[7]. Reduction of electron energy while maintaining atomic resolution will drastically reduce radiation damage in delicate organic and biological samples[8]. Undoubtedly, this revolutionary technology will impact many areas of science and technology, including physics, chemistry, materials science, geology, archeology, biology, medicine, manufacturing, etcetera. However, significant unresolved issues still remain. Recent experience with TEM shows that the optimum corrected state can be maintained for only a few minutes, after which the microscope drifts away and must be re-adjusted[9-11], a serious concern to microscope designers and users alike.

Here we discuss how resolution depends on the degree to which aberrations are corrected: resolution is exquisitely sensitive to small deviations from full correction, and is intrinsically unstable against small fluctuations. For instance, to achieve at least 90% of the resolution improvement afforded by correction of the 3$^{rd}$ order spherical aberration coefficient $C_3$, with simultaneous correction of the chromatic aberration coefficient $C_c$, $C_3$ must be corrected to within 1/10,000$^{th}$ of its uncorrected value. For a typical TEM with $C_3 = 1$ mm, correction must therefore be accurate and stable to within 0.1 μm. Correction of the 5$^{th}$ order spherical aberration $C_5$, in addition to $C_c$ and $C_3$, is even harder, and it appears unlikely that a stable state could be maintained for any significant length of time. Aberration correction may utilize either axially symmetric electron mirrors[12] as in LEEM/PEEM, or sophisticated multipole optics[13] as in (Scanning) TEM. That such aberration-corrected TEM instruments have stringent environmental and electronic stability requirements is well documented[14]. However, the fact that corrected electron optical instruments are *intrinsically unstable* does not appear to be widely recognized or appreciated.

In the simplest approach we define the resolution, δ, as follows:

$$\delta^2 = (0.61\lambda/\alpha)^2 + (C_c\varepsilon\alpha)^2 + (C_3\alpha^3)^2 + (C_5\alpha^5)^2 + .... \tag{1}$$

where $\lambda$ is the electron wavelength, and $\varepsilon$ the normalized energy spread $\Delta E/E$. The first term, $0.61\lambda/\alpha$, is the Rayleigh limit, due to a contrast aperture with angular range +/- $\alpha$. The best resolution occurs when $d\delta/d\alpha = 0$. We consider three limiting cases in which two aberration coefficients are set to zero, and the third is free to vary, leading to the following power-laws:

$$
\begin{aligned}
&C_3 = C_5 = 0: \quad \delta \propto (C_c)^{1/2} \\
&C_c = C_5 = 0: \quad \delta \propto (C_3)^{1/4} \\
&C_c = C_3 = 0: \quad \delta \propto (C_5)^{1/6}
\end{aligned}
\tag{2a-c}
$$

The same dependencies are obtained from a wave-optical theory based on the Contrast Transfer Function (CTF), which quantifies the aberrations of the objective lens. The CTF is given by[15]:

$$W = e^{i2\pi\chi} = \cos(2\pi\chi) + i\sin(2\pi\chi)$$
$$\chi = \frac{1}{2}C_1\lambda q^2 + \frac{1}{4}C_3\lambda^3 q^4 + \frac{1}{6}C_5\lambda^5 q^6 + ... \tag{3a,b}$$

$C_1$ is the defocus, and $q$ the spatial frequency. The point resolution (i.e. the value of $q$ at the first zero crossing of $W$, in nm$^{-1}$) is given by $Im(W) = 0$ when relative phase shifts in the exit wave function are near-zero (weak-phase object). For a strong-phase object (relative phase shifts around π) or amplitude object (exit wave dominated by structure factor contrast) it is given by[16] $Re(W) = 0$. With $C_1 = C_5 = 0$ we obtain for weak-phase (eq. 4a) and strong-phase/amplitude (eq. 4b) objects:

$$\sin(\frac{\pi}{2}C_3\lambda^3 q_r^4) = 0; \quad \frac{\pi}{2}C_3\lambda^3 q_r^4 = \pi \tag{4a}$$

$$\cos(\frac{\pi}{2}C_3\lambda^3 q_r^4) = 0; \quad \frac{\pi}{2}C_3\lambda^3 q_r^4 = \frac{\pi}{2} \tag{4b}$$

i.e. $\delta = 1/q_r \propto C_3^{1/4}$, the same as eq. (2b). Similarly, when $C_1 = C_3 = 0$, $\delta = 1/q_r \propto C_5^{1/6}$, as in eq. (2c). Chromatic aberrations are captured in the envelope function[15]:

$$E_c(q) = \exp(-\frac{(\pi C_c \lambda q^2)^2}{16\ln 2}\varepsilon^2) \tag{5}$$

Taking $E_c(q_i) = e^{-2}$ to define the information limit[15], the $C_c$ limited resolution is given by:

$$C_c q_i^2 = \frac{4\sqrt{2\ln(2)}}{\pi\lambda\varepsilon} \tag{6a-b}$$

$$\delta = 1/q_i \propto C_c^{1/2}$$

as in eq. 2a. Thus, eqs.1 and 3 lead to identical results.

Looking at the aberration coefficients individually, with the other coefficients set to zero, for the resolution to reach 50% (10%) of its uncorrected value, $C_c$ must be corrected to better than 25% (1%), $C_3$ to better than 6.25% (0.01%), and $C_5$ to better than 1.5% (1 ppm); the window in which the benefits of aberration correction can be obtained shrinks rapidly with increasing order. The stability of the corrected state is determined by the derivatives of resolution with respect to the aberration coefficients. When these derivatives are zero, the system is stable and protected from small fluctuations. However, these derivatives scale with $C_c^{-1/2}, C_3^{-3/4}, C_5^{-5/6}$, diverging as the corrected state is approached. That is, the corrected state is intrinsically unstable, and the more fully it is realized, the more unstable it is.

In the following we use a more realistic and complete scheme of calculating resolution. Using the CTF to calculate images for specific objects[16], all aberration coefficients up to 5$^{th}$ order are set at the actual values calculated for a $C_c / C_3$ corrected LEEM/PEEM instrument[17]. We calculate images at $C_1 = 0$ for amplitude objects[18] as commonly encountered in LEEM, and extract the resolution. We use $\Delta E_0 = 0.25$ eV, and the column energy $E = 15$ keV. Figure 1a shows resolution vs. $C_3$ (normalized to the uncorrected value) with $C_c$ ranging from uncorrected (100%) to fully corrected (0%). As $C_c$ decreases a deep cusp develops near $C_3 = 0$. The minimum does not reach zero, as higher order coefficients[17] (such as $C_{cc}$, and $C_5$) are set at the non-zero values obtained from raytracing. The minimum is shifted to a slightly negative value of $C_3$, offsetting the positive value of $C_5$. The dotted $\delta \sim C_3^{1/4}$ line is in close agreement with the full calculations when $C_c = 0$. These results do not depend significantly on $E_0$. The effects of non-zero defocus will be discussed in more detail below.

In figure 1b $C_3 = 0$ and $C_c$ is varied for different values of $E_0$. The dotted line shows $\delta \propto C_c^{1/2}$. For all values of $E_0$, as $C_c$ increases the simulations follow $\delta \propto C_c^{1/2}$ closely. Figure 1c compares $\Delta E_0 = 0.25$ eV (cold field emission[19]) with $\Delta E_0 = 0.75$ eV (typical LaB$_6$ gun[20]). The ordinate is not normalized, to highlight differences on an absolute scale. Dotted lines are individually scaled $\delta \sim C_c^{1/2}$ lines. While near $C_c = 0$ the two cases are almost identical, for the uncorrected situation the difference is significant. As expected, the minimum is steeper and narrower as $\Delta E_0$ increases and chromatic aberration is more significant. Very similar results were obtained for weak-phase objects[18], or by plotting point resolution (eq. (4)) vs. $C_3$ for amplitude and weak-phase objects.

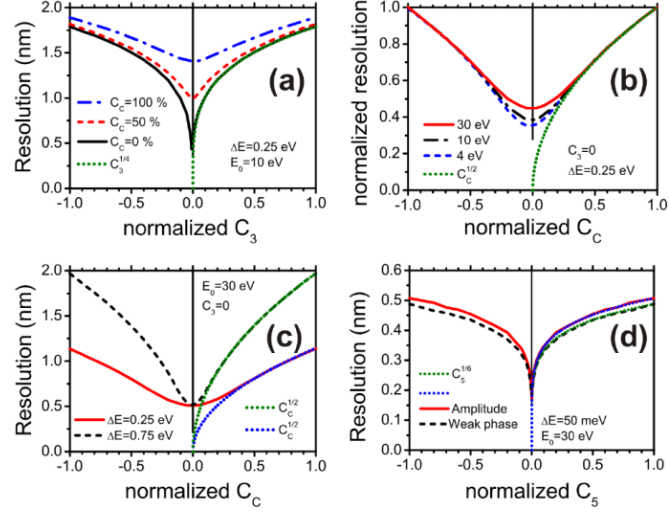

Figure 1. (a) Resolution of an amplitude object vs. the normalized value of $C_3$, for different settings of $C_c$ (uncorrected = 100%, fully corrected = 0%). Start energy $E_0 = 10$ eV, $\Delta E_0 = 0.25$ eV. Green dotted line: $C_3^{1/4}$ prediction of eqs. (2b) and (4b). (b) Resolution (normalized to uncorrected values) vs. normalized value of $C_c$, with $C_3 = 0$, for $E_0 = 4$, 10 and 30 eV.(c) Resolution vs. normalized value of $C_c$, with $C_3 = 0$, for $\Delta E_0 = 0.25$ and 0.75 eV. Dotted lines in (b) and (c): $C_c^{1/2}$ prediction of eqs. (2a) and (6b). (d) Resolution vs $C_5$ for $C_c / C_3$ corrected LEEM with $\Delta E_0 = 50$ meV. The microscope has (near) atomic resolution of 0.17 nm. However, this corrected state is very fragile: a 0.003 excursion from the minimum along the abscissa degrades the resolution by 20%. The dotted lines show the $C_5^{1/6}$ prediction.

Finally, in figure 1d we consider a 'super-corrected' LEEM in which $C_c = C_3 = 0$. The remaining chromatic aberrations are minimized by an energy-filtered gun with $\Delta E_0 = 50$ meV. The cusp around $C_5 = 0$ shows the predicted $C_5^{1/6}$ dependence. This microscope promises atomic resolution (0.17 nm) with 30 eV electrons, limited by the higher order chromatic terms. It is conceivable that such an instrument be designed and built, using an electron mirror with at least four electrostatic elements to control $C_1$, $C_3$, $C_5$ and $C_c$. $C_5$ must be reduced from 14.5 m to < 0.5 mm[17], $C_3$ from ~300 mm to ~ -10 μm, with a stability of ~0.5 μm, and $C_1$ must be controlled to better than 2 nm. However, the corrected state would be extremely fragile due to the very steep and narrow cusp in figure 1d.

Turning to transmission microscopes, the $C_3$-limited TEM resolution at Scherzer defocus is given[15] by $\delta = 0.66 C_3^{1/4} \lambda^{3/4}$, the same $C_3^{1/4}$ dependence as seen before. In the following we focus on the region of slightly negative $C_3$ and slightly positive $C_1$ which previous studies have shown to give the highest resolution imaging results. Figure 2a shows the point resolution (the $q$ value at the first zero-crossing of $\sin(2\pi\chi)$, in nm$^{-1}$) vs $C_3$ for a weak-phase object in a TEAM-like microscope[2] ($C_5 = 5$ mm, 300 keV). We note the presence of a narrow, ridge-shaped optimum-resolution band diagonally across the figure, with optimum performance along the yellow line near the center. The CTF along this line is optimally balanced over all spatial frequencies below the point resolution and is characterized by a single parameter, $0 < \phi < \pi/2$ (see supplementary material). At $\phi = \pi/2$ figure 2a shows a singularity where two line-shaped singularities intersect.

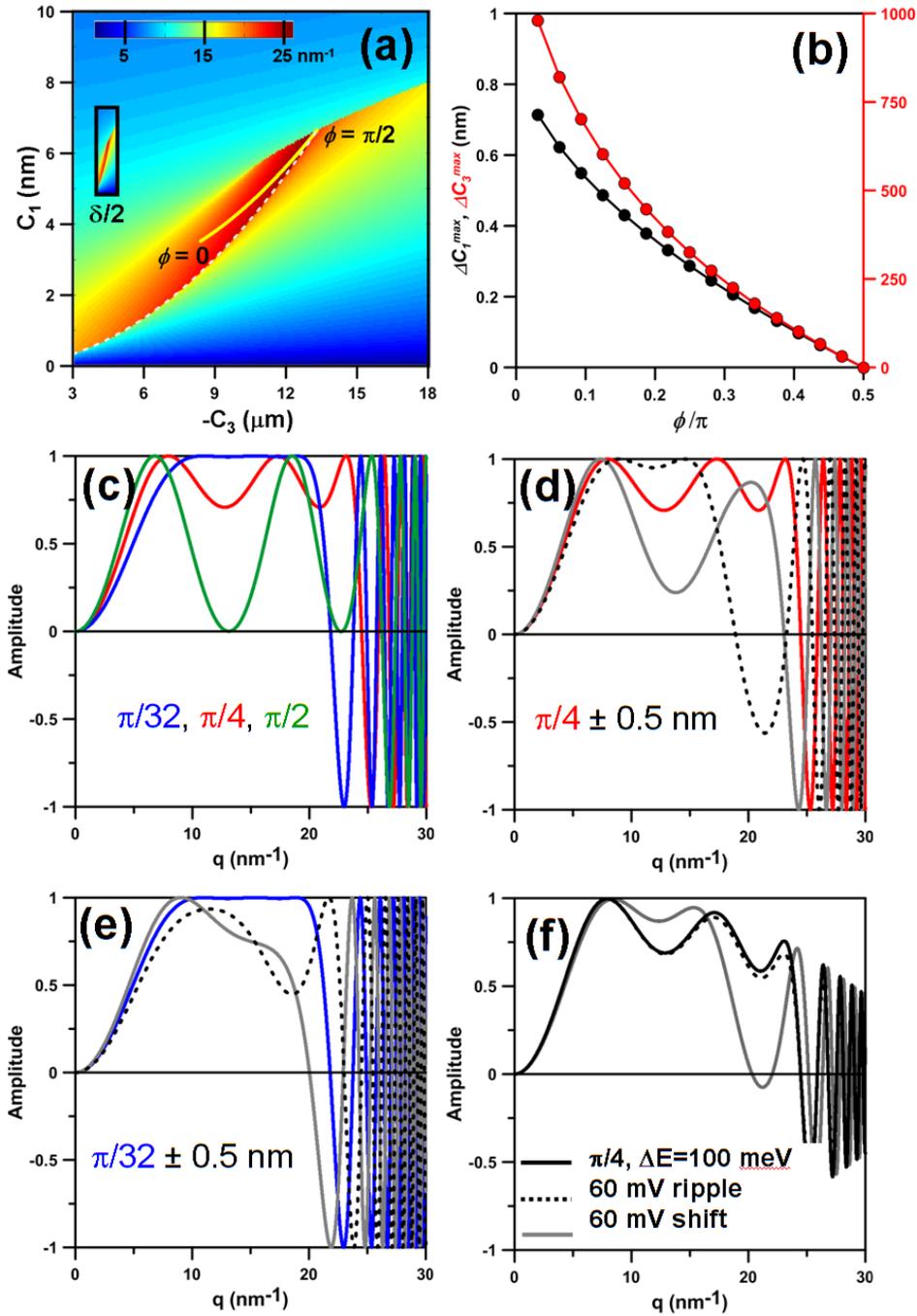

Figure 2. (a) (a) Point resolution (in nm$^{-1}$) vs $C_1$ and $C_3$ near $C_3 = 0$. Yellow line: optimized performance as a function of $\phi$, ranging from $\phi = 0$ to $\pi/2$. White dashed line: abrupt instability in the transfer function. The inset below the scale bar, labeled $\delta/2$, shows the relative size of this map if resolution is improved by a factor 2 by reduction of $C_5$. (b) Instability budgets for $C_1$ and $C_3$, defined by the distance between the yellow and white-dashed lines in (a), as a function of $\phi$. (c) CTF for different values of $\phi$. (d) CTF at $\phi = \pi/4$ with negative (dashed grey) and positive (solid grey) 0.5 nm additional defocus. (e) as (d) for $\phi = \pi/32$. The sensitivity for a small defocus is greater in (d) than in (e), in agreement (b). (f) CTF at $\phi = \pi/4$ with $\Delta E = 100$ meV (black line). Dashed line: additional high voltage ripple $v = 60$ mV. Solid grey line: high voltage shift of -60 mV.

To the right of this point the resolution is always inferior. Along the white dashed line the CTF becomes unstable and the resolution drops abruptly. The distance between the yellow and white lines is the largest deviation that can be tolerated without a significant loss of resolution, defining an 'instability budget' for $C_1$ and $C_3$ (see supplementary material). Figure 2b plots the instability budgets as a function of $\phi$. The budget for $C_1$ decreases from ~0.7 nm at $\phi = \pi/32$ to 0 nm at $\phi = \pi/2$, while for $C_3$ it changes from ~1 μm to 0 μm. At the same time, δ changes from 46 pm at $\phi = \pi/32$ to 38 pm at $\phi = \pi/2$. Note that these instability budgets are not fixed: they depend on the value of $\phi$ selected by the operator. Figure 2c shows the CTF for $\phi = \pi/32$, $\pi/4$, and $\pi/2$. For $\phi = \pi/4$ (Scherzer defocus) the instability budgets for $C_1$ and $C_3$ are ~0.28 nm and ~0.32 μm, respectively. For $\phi = \pi/32$ resolution is somewhat worse, but stability has improved. In figures 2d and 2e we plot the CTF at $\phi = \pi/4$ (2d) and $\phi = \pi/32$ (2e), with additional offsets in $C_1$ of ±0.5 nm, exceeding the instability budget for $\phi = \pi/4$, but well below it for $\phi = \pi/32$. In figure 2d the CTF is strongly affected, with a deep minimum at $q \approx 20$ nm$^{-1}$ for -0.5 nm defocus. The CTF in figure 2e is much less affected, with a point resolution well above 20 nm$^{-1}$ at -0.5 nm defocus, and 20 nm$^{-1}$ at +0.5 nm defocus. This may seem counter-intuitive. To obtain 50 pm resolution, it would appear that the CTF at $\phi = \pi/4$ is better than at $\phi = \pi/32$, as it has a higher point resolution. However, it is also significantly less stable. In practice, one may prefer the small loss in resolution at $\phi = \pi/32$, as it provides a better instability budget. The map in figure 2a is not specific for a TEAM-like instrument. Every electron microscope where $C_1$ and $C_3$ can be adjusted for a given $C_5$ behaves in the same manner. As shown in the supplement, resolution scales with $C_5^{1/6}$, while the instability budgets for $C_1$ and $C_3$ scale with $C_5^{1/3}$ and $C_5^{2/3}$, respectively. The advent of $C_3$ correction led to an improvement in resolution by about a factor 2. Could we gain another factor 2 by reducing $C_5$? The small inset labeled $\delta/2$ in figure 2a shows the relative size of the resolution map resulting from a reduction of $C_5$ from 5 mm to 0.08 mm, required to improve the resolution by a factor 2. Regardless of the fact that this would present huge, possibly insurmountable challenges in controlling numerous other aberrations, it is clear from the diminutive size of this map that the leading aberrations $C_1$ and $C_3$ could not be controlled with sufficient accuracy and stability to make such an improvement possible; the instability budgets have shrunken to near-nothing.

When $C_c$ is corrected the system is -to first order- insensitive to small fluctuations in the electron gun potential. But when only $C_3$ is corrected, a small shift $v$ in the electron gun potential V is equivalent to a focus shift $\Delta C_1 = C_c \cdot v/V$. To appreciate the difference between a high frequency ripple, vs. a static shift of the gun potential, we refer to figure 2f. We use $C_c = 1.6$ mm, an energy spread $\Delta E = 100$ meV, and no high voltage ripple (solid black line). The slow drop-off for $q > 10$ nm$^{-1}$ is due to the chromatic envelope function, eq. (5). The dashed line results when -in addition to the energy spread of 100 meV- we introduce an additional high voltage ripple of 60 mV. The effect is minor: a slightly stronger drop-off at higher q-values. In contrast, the grey line uses $\Delta E = 100$ meV, no HV ripple, plus a *static* HV shift of -60 mV, equivalent to $\Delta C_1 = -0.32$ nm. Now the CTF has changed dramatically, and the CTF amplitude at $q = 20$ nm$^{-1}$, critical for a spatial resolution of 50 pm, has dropped to about zero. To keep $\Delta C_1$ stable to within 0.16 nm, the absolute voltage stability (i.e. immunity against drift) must be better than 30 meV at 300 keV (10 meV at 100 keV), a relative stability of 0.1 ppm. When $C_c$ is not corrected, using a gun monochromator[21] reduces the energy spread *prior* to acceleration to the final beam energy. However, instabilities in the acceleration stage (i.e. drift in the high tension supply for the electron gun) remain unremedied. A similar effect is caused by instabilities in the objective lens current, I: a small current shift $i$ causes a defocus shift $\Delta C_1 = 2C_c \cdot i/I$. Thus, the objective lens power supply must have a relative stability of $5 \times 10^{-8}$. Such extraordinary long-term drift stabilities are extremely difficult to realize experimentally. For a TEAM-like instrument, the gun high voltage can drift over 100 mV on a timescale from minutes to hours, depending on the quality of the air conditioning in the room[22]. Of course, mechanical drift of the sample along the beam direction, as well as undulations in the thin sample foil also give rise to defocus shifts.

These results are not limited to LEEM or TEM, but hold for any electron optical instrument. As more aberration coefficients are corrected, the widths of the cusps within which correction must be maintained become increasingly narrow. Our findings shed new light on the short-lived corrected state observed in state-of-the-art aberration corrected TEM instruments[9-11]. With resolution exquisitely sensitive to the residual values of the aberration coefficients, even minute mechanical and electronic drifts are strongly amplified. Uncorrected chromatic aberrations can create a small 'island of stability' around the corrected state. Figure 1a shows that this island is reasonably broad when $C_c$ is uncorrected. But when $C_c$ is corrected it shrinks dramatically, leaving the improved corrected state much less protected. Additionally, as the quality of the corrected state is improved it becomes increasingly difficult to reliably put and keep the system into that state. To put the CTF at $\phi = \pi/4$ within 30% of the instability budget, we must measure $C_1$ and $C_3$ with sufficient accuracy, and know the value of $C_5$ to better than 1.5%. When we reduce $C_5$, increase in instability outstrips improvement in resolution, posing a fundamental limit on the resolution that is ultimately achievable.

While the intrinsic instability identified in this paper presents a serious challenge, it may be possible to monitor the state of the microscope in real time, and adjust the instrument settings 'on the fly' to maintain the corrected condition, much like most aircraft flying today are inherently unstable without sophisticated electronic control systems. Identifying suitable measurable parameters to fully quantify the state of the microscope during routine sample observation is a task that presently remains unresolved.

**Acknowledgements:** The authors thank Phil Batson for useful discussions. We also thank the referees for their helpful comments. This research was funded in part by the Netherlands Organization for Scientific Research (NWO) via an NWO-Groot Grant ('ESCHER').